\documentclass[preprint,showpacs,pre,amsmath,amssymb,endfloats*]{revtex4}
\usepackage{graphicx}
\usepackage{dcolumn}
\begin{document}

\preprint{EK8907}

\title{Evolution of the isotropic to nematic phase transition in octyloxycyanobiphenyl+aerosil dispersions}

\author{A. Roshi and G. S. Iannacchione}
\affiliation{Department of Physics, Worcester Polytechnic
Institute, Worcester, Massachusetts 01609, USA}

\author{P. S. Clegg and R. J. Birgeneau}
\affiliation{Department of Physics, University of Toronto,
Toronto, Ontario M5S 1A7, Canada}

\date{\today}


\begin{abstract}
High-resolution ac-calorimetry has been carried out on dispersions
of aerosils in the liquid crystal octyloxycyanobiphenyl ($8OCB$)
as a function of aerosil concentration and temperature spanning
the crystal to isotropic phases. The liquid-crystal $8OCB$ is
elastically stiffer than the previously well studied
octylcyanobiphenyl ($8CB$)+aerosil system and so, general quenched
random disorder effects and liquid-crystal specific effects can be
distinguished. A double heat capacity feature is observed at the
isotropic to nematic phase transition with an aerosil independent
overlap of the heat capacity wings far from the transition and
having a non-monotonic variation of the transition temperature. A
crossover between low and high aerosil density behavior is
observed for $8OCB$+aerosil. These features are generally
consistent with those on the $8CB$+aerosil system. Differences
between these two systems in the magnitude of the transition
temperature shifts, heat capacity suppression, and crossover
aerosil density between the two regimes of behavior indicate a
liquid crystal specific effect. The low aerosil density regime is
apparently more orientationally disordered than the high aerosil
density regime, which is more translationally disordered. An
interpretation of these results based on a temperature dependent
disorder strength is discussed. Finally, a detailed thermal
hysteresis study has found that crystallization of a well
homogenized sample perturbs and increases the disorder for low
aerosil density samples but does not influence high density
samples.
\end{abstract}

\pacs{64.70.Md, 61.30.Eb, 65.40.Ba}


\maketitle


\section{INTRODUCTION}
\label{sec:intro}

The effect of quenched random disorder on phase structure and
transitions is an important area of study that continues to
attract a great deal of research. Disorder is ubiquitous (ideal
pure transitions being the exception rather than the rule in
nature) and the effect on phase transitions can be profound. Phase
transitions are modified depending on the aspect of the system
affected by the disorder, on the dimensionality, and on the number
of components to the order parameter. Of particular interest is a
linear coupling between quenched random disorder and the order
parameter, which allows for a random-field theoretical approach.
Also, the order of the transition is crucial as first-order
transitions have additional considerations compared to continuous
transitions. This is due to the presence of two-phase coexistence
(hence interfaces between ordered and disordered regions),
intrinsically finite correlation length at the transition, and
hysteresis effects for first-order compared to continuous phase
transitions. This has made the experimental and theoretical
studies of quenched random disorder effects at first-order
transitions challenging.

Liquid-crystals (LC) are a particularly attractive system for the
study of phase transitions into partially ordered phases. This
makes them especially interesting for the study of the effects of
quenched random disorder (QRD), which are typically introduced by
the random fixed dispersion of solid surfaces. In LC+aerosil
systems, the quenched random disorder is created by a dispersed
gel of aerosil particles and is varied by changing the density of
aerosils in the dispersion. A convenient measure of the introduced
disorder is the grams of silica per cm$^3$ of liquid-crystal,
denoted the conjugate silica density $\rho_S$, which is directly
related to the surface area of solids as well as the mean-distance
between solid surfaces \cite{Germano98,Germano03}. The aerosils
used are silica spheres that can hydrogen bond together to form a
fractal-like random gel. Studies have previously been carried out
by various groups on liquid crystals in an aerogel medium
\cite{Bellini01}. Aerogels are self-supporting structures and this
places a lower limit on the disorder strength that can be probed.
By contrast, the aerosil gel provides a weaker and more easily
controlled perturbation, and thus opens up a physically
interesting regime.

In this work, we study the effect of quenched random disorder due
to a dispersed thixotropic aerosil gel on the weakly first-order
isotropic to nematic (\textit{I}-\textit{N}) phase transition. The
calorimetric results for the nematic to smectic-\textit{A} phase
transition in $8OCB$+aerosil samples have been previously reported
and were shown to be consistent with results in $8CB$+aerosil
samples \cite{Clegg03}. At the \textit{I}-\textit{N} transition,
the orientational order has a finite correlation length and is
established in three dimensions, which is describable by a
symmetric and traceless $2nd$ rank tensor $Q_{ij}$ \cite{Gennes93}
(as such, it possesses only five independent components). Thus,
nematic order belongs, in principle, to a $d = 3$, $n = 5$
Heisenberg class. However, by ignoring any biaxial character and
aligning the orientation axis with a principle axis of a local
frame, this tensor can be split into a scalar order parameter
($S$) measuring the magnitude of orientational order about the
orientation axis and a ''headless'' vector called the nematic
director $\hat{n}$ ($\hat{n} = -\hat{n}$) describing the spatial
orientation of this axis. In this simplified view, nematic order
is described on short length scales by $S$ and on longer length
scales by $\hat{n}$, which is useful in describing the elastic
properties of the nematic structure. These measures of nematic
order are related to the quadrupolar nematic order parameter by
$Q_{ij} = \frac{1}{2} S (3\hat{n}_i\hat{n}_j - \delta_{ij})$.

In principle, the effect of the aerosil gel network on the
orientational order of the nematic phase is two fold. The silica
gel firstly dilutes the liquid crystal and secondly creates a
preferred local orientation \cite{Gingras,Feldman00}. In addition,
the first-order transition from the isotropic to the nematic phase
necessitates the formation of interfaces between coexisting
domains/phases, which must occur within the available void spaces.
The latter effect is the classic result of quenched random
disorder; a distribution of transition temperatures due to the
nucleation of ordered domains within voids having some size
distribution. This leads to short-range order (SRO), a rounding of
the transition, and suppression of the first-order character of
the transition \cite{Imry79}. The contribution of a random
preferred local orientation effect to the total hamiltonian can be
represented as
\begin{equation}
  \label{eq:Hr}
  \mathcal{H}_{rf} = -\sum_i g_2 (\bf{h}_i \cdot \bf{\hat n}_i)^2 \;
\end{equation}
where $\bf{\hat n}_i$ is the orientation of the molecules over
some small region where the orientation is approximately constant
and $\bf{h}_i$ is the random influence of the silica surface. The
variance of this random field $\langle h^2\rangle$ should be
proportional to the density of solids dispersed in the LC medium.
This term is squared due to the effective inversion symmetry of
the molecules in the nematic phase. Since the nematic order
parameter is quadratic in $\bf{\hat n}_i$ due to the same
inversion symmetry \cite{Gennes93}, Eq.~(\ref{eq:Hr}) is also
linear in the order parameter and hence constitutes a random field
(RF) interaction. Recently, Eq.~(\ref{eq:Hr}) has also been
interpreted as a random-anisotropy (RA) interaction
\cite{Feldman00} but this seems only applicable to systems
describable by a pure vector $n = 3$ order parameter. The
formation of interfaces and the resulting surface energy penalties
places restrictions on the effects of $\langle h^2\rangle$
depending on the elasticity of the nematic. Light scattering
measurements have shown that the nematic phase in liquid crystal
and aerosil dispersions breaks up into large (micron size) but
finite-size domains \cite{Bellini98}. In addition, more extensive
optical studies focussing on the nature of the nematic director
structure well below the \textit{I}-\textit{N} transition have
shown that the director correlation length $\xi_{\hat n}$ decays
exponentially with distance, which is a hallmark of short-range
order \cite{Bellini00,Bellini02}. These features are consistent
with an RF interaction for nematics with QRD.

To date, the most thoroughly studied LC+aerosil system is the
dispersion of type-300 aerosil in octylcyanobiphenyl ($8CB$),
denoted $8CB$+aerosil. Detailed calorimetric
\cite{Zhou97b,Germano98,Marinelli01}, x-ray scattering
\cite{Park02,Leheny03}, x-ray intensity fluctuation spectroscopy
\cite{Retsch02}, static and dynamic light-scattering
\cite{Bellini98,Bellini00,Bellini02}, and deuterium NMR
\cite{Jin01} studies on the nematic to smectic-\textit{A}
(\textit{N}-Sm\textit{A}) and the isotropic to nematic
(\textit{I}-\textit{N}) phase transitions of this system have
shown that there are clear quenched random-field characteristics
as well as finite-size scaling effects \cite{Germano03}.

Calorimetry measurements on $8CB$+aerosil samples have been
particularly useful in yielding detailed information on both the
\textit{I}-\textit{N} and the \textit{N}-Sm\textit{A} phase
transitions \cite{Germano98}. The results for both transitions
show a complex dependence of the transition temperature on the
aerosil density. While the \textit{N}-Sm\textit{A} heat capacity
peak remains sharp and evolves towards 3D-\textit{XY} behavior
with increasing silica density, the \textit{I}-\textit{N} behavior
is more complicated. For silica densities below $\rho_S \sim
0.1$~g~cm$^{-3}$, two heat capacity peaks, closely spaced in
temperature, were observed. At higher aerosil densities, the heat
capacity peaks for both the \textit{I}-\textit{N} and the
\textit{N}-Sm\textit{A} transitions displayed a highly smeared and
non-singular features. Deuterium NMR measurements on deuterated
$8CB$+aerosil dispersions, which were carried out over a wide
range of silica densities, showed that the magnitude of the
orientational order $S$ below the \textit{I}-\textit{N} transition
temperature was essentially unchanged from bulk behavior
\cite{Jin01}. The amount of liquid crystal reorientation for
field-cooled samples upon rotation within the DNMR field is small
and decreases continuously with silica density up to $\rho_S =
0.094$~g~cm$^{-3}$ (the units will be dropped hereafter)
confirming distinct low and high $\rho_S$ behavior. A x-ray
intensity fluctuation spectroscopy (XIFS) study have found
evidence of aerosil gel dynamics in $8CB$+aerosil dispersions
indicating an elastic coupling between the gel and LC
\cite{Retsch02}. The optical, calorimetric, DNMR, and XIFS results
all appear to be consistent with a model in which director
fluctuations are suppressed with increasing aerosil density.

The present work focusses on a different liquid crystal --
octyloxycyanobiphenyl ($8OCB$) -- having dispersed in it the same
type of aerosil over a comparable range of silica densities as the
well-studied $8CB$+aerosil system. This liquid-crystal has several
important differences from the closely related $8CB$. The
liquid-crystal $8OCB$ has stronger smectic and nematic
interactions than $8CB$ as evidenced by the higher transition
temperatures, the larger bare correlation lengths for smectic
interactions \cite{Litster79}, and the larger elastic constants.
More specifically, $8OCB$ has a $17\%$ larger bend, $36\%$ larger
twist, and $10\%$ larger splay nematic elastic constants than
$8CB$ (in the single elastic constant approximation, $8OCB$ has a
$\approx 20\%$ greater $K_N$ than $8CB$ with an overall
uncertainty of $5\%$) \cite{Bradshaw85}. Thus, comparison of
behaviors between $8CB$+aerosil and $8OCB$+aerosil systems allow
for the isolation of general quenched random-disorder (QRD)
effects from material specific effects, in this case the
elasticity of the liquid-crystal host medium.

In general, this work reveals a non-monotonic silica density
dependence of the \textit{I}-\textit{N} and
\textit{N}-Sm\textit{A} transition temperatures similar to that
observed for $8CB$+aerosil but occurring over a larger $\rho_S$
range for $8OCB$+aerosil. The calorimetric results presented here
for the \textit{I}-\textit{N} transition reveal the onset of a
double transition peak for $\rho_S > 0.1$ with a $\rho_S$
dependence on the temperature distance between the two heat
capacity peaks. Evidence is presented that the first-order
character of the \textit{I}-\textit{N} transition continuously
decreases with silica content, becoming approximately zero for
$\rho_S \gtrsim 0.7$. Over the entire range of $\rho_S$ studied
here, the heat capacity temperature dependence away from the
immediate vicinity of the transition region is bulk-like and
independent of silica content.

We speculate that the variance of the disorder $\langle
h^2\rangle$ may change through a first-order transition for
nematics to account for these observations. Such a variation of
the disorder strength may be due to the silica surfaces
introducing a low-order, paranematic-like, boundary layer
initially screening the remaining liquid crystal material. The
thickness of this boundary layer is strongly temperature dependent
in the immediate vicinity of the \textit{I}-\textit{N} transition
and as it shrinks, the screening becomes weaker.

Section~\ref{sec:exp} describes the preparation of the
$8OCB$+aerosil dispersions as well as the ac-calorimetry technique
employed. Given in Section~\ref{sec:results} is a presentation of
the results. All results are then discussed in
Section~\ref{sec:disc} and related to results from previous
LC+aerosil studies. Directions for future study will also be
discussed.


\section{EXPERIMENTAL TECHNIQUES}
\label{sec:exp}

The liquid crystal $8OCB$, purchased from Aldrich, was used after
degassing in the isotropic phase for $1$ hour. This liquid-crystal
molecule has an aliphatic tail attached by an oxygen link to the
rigid biphenyl core and a polar cyano head group ( $M_w =
307.44$~g~mol$^{-1}$ ). This oxygen link constitutes the sole
molecular difference between $8OCB$ and $8CB$. Pure $8OCB$ has a
weakly first-order isotropic to nematic transition at $T_{IN}^o
\sim 353$~K and a second-order nematic to smectic-\textit{A}
transition at $T_{NA}^o \sim 340$~K. At lower temperatures, the
strongly first-order Crystal-SmA transition occurs reproducibly on
heating at $T_{CrA}^o \sim 328$~K and, as usual, can be greatly
supercooled.

The hydrophilic type-300 aerosil obtained from Degussa
\cite{Degussa} was thoroughly dried at $\sim 300^o\textrm{C}$
under vacuum for a couple of hours prior to use. The hydrophilic
nature of the aerosils arises from the hydroxyl groups covering
the surface and allows the aerosil particles to hydrogen bond to
each other. This type of bonding is weak and can be broken and
reformed, which leads to the thixotropic nature of gels formed by
aerosils in an organic solvent. Crystallization severely disrupts
the gel. The specific surface area measured by the manufacturer
via BET nitrogen isotherms is $300$~m$^2$~g$^{-1}$ and each
aerosil sphere is roughly $7$-nm in diameter. However, SAXS
studies have shown that the basic aerosil unit consists of a few
of these spheres fused together during the manufacturing process
\cite{Germano98}. Each $8OCB$+aerosil sample was created by mixing
appropriate quantities of liquid crystal and aerosil together,
then dissolving the resulting mixture in spectroscopic grade (low
water content) acetone. The resulting solution was then dispersed
using an ultrasonic bath for about an hour. As the acetone
evaporates from the mixture, a fractal-like gel forms through
diffusion-limited aggregation. Small angle x-ray studies have
shown that the aerosil gel dispersion has a fractal structure and
no preferred orientation \cite{Germano98} on the micron-long
length scales of nematic order \cite{Bellini00}.

At room temperature, $8OCB$ is a crystalline solid even in the
presence of high aerosil density. Care was taken to avoid
crystallization of $8OCB$ and possible damage to the aerosil gel,
especially for low silica densities. For this calorimetry study,
the mixture after slow solvent evaporation was allowed to
crystallize, and the solid sample was transferred into the
calorimetry cell. The cell was then sealed, the heater and
thermometer attached, and the cell was heated into the isotropic
phase. The sample was then remixed by placing the assembly in an
ultrasonic bath for over an hour. The cell and sample temperature
was kept elevated during the mounting of the sealed cell into the
calorimeter by maintaining current through the heater. This sample
preparation protocol also allows a controlled entry into the
crystal phase. However, since the cell is sealed, the in situ
remix could not be inspected and so some small dispersion
inhomogeneity may remain.

High-resolution ac calorimetry was performed using two home-built
calorimeters at WPI. The sample cell consisted of a silver
crimped-sealed envelope $\sim 10$~mm long, $\sim 5$~mm wide, and
$\sim 0.5$~mm thick (closely matching the dimensions of the
heater). After the sample was introduced into a cell having an
attached $120$-$\Omega$ strain-gauge heater and $1$-M$\Omega$
carbon-flake thermistor, a constant current was placed across the
heater to maintain the cell temperature well above $T_{IN}$. The
filled cell was then placed in an ultrasonic bath to remix the
sample. After remixing, the cell was mounted in the calorimeter,
the details of which have been described elsewhere \cite{Yao98}.
In the ac-mode, power is input to the cell as $P_{ac} e^{i\omega
t}$ resulting in temperature oscillations with amplitude $T_{ac}$
and a relative phase shift of $\varphi \equiv \Phi + \pi/2$, where
$\Phi$ is the absolute phase shift between $T_{ac}(\omega)$ and
the input power. The specific heat at a heating frequency $\omega$
is given by
\begin{equation}
  \label{eq:ReCp}
  C_p = \frac{[C_{filled}' - C_{empty}]}{m_{sample}} = \frac{P_{ac}cos\varphi / \omega
  |T_{ac}| - C_{empty}}{m_{sample}} \;,
\end{equation}
\begin{equation}
  \label{eq:ImC}
  C_{filled}'' = \frac{P_{ac}}{\omega |T_{ac}|} sin\varphi -
  \frac{1}{\omega R} \;,
\end{equation}
where $C_{filled}'$ and $C_{filled}''$ are the real and imaginary
components of the heat capacity, $C_{empty}$ is the heat capacity
of the cell and silica, $m_{sample}$ is the mass in grams of the
liquid crystal (the total mass of the $8OCB$+aerosil sample was
$\sim 20$~mg, which yielded $m_{sample}$ values in the range of
$13-20$~mg), and $R$ is the thermal resistance between the cell
and the bath (here, $\sim 200$~K~W$^{-1}$).
Equations~(\ref{eq:ReCp}) and (\ref{eq:ImC}) require a small
correction to account for the finite internal thermal resistance
compared to $R$, and this was applied to all samples studied here
\cite{Germano97}. Measurements were conducted at various
frequencies in order to ensure the applicability of
Eqs.~(\ref{eq:ReCp}) and (\ref{eq:ImC}) by checking that
$C_{filled}^{''} \approx 0$ through the effective
\textit{N}-Sm\textit{A} transition at $T^\ast$ and that $C_p$ was
independent of $\omega$. All data presented here were taken at
$\omega = 0.1473$~s$^{-1}$ at a scanning rate of less than $\pm
100$~mK~h$^{-1}$, which yield essentially static $C_p$ results.
All $8OCB$+aerosil samples experienced the same thermal history
after mounting; six hours in the isotropic phase to ensure
homogeneous gelation, then a slow cool deep into the smectic phase
before beginning the first detailed scan upon heating.


\section{RESULTS}
\label{sec:results}

\subsection{General description}

The heat capacity of the pure $8OCB$ liquid-crystal is in good
agreement with previously published results
\cite{Kasting80,Garland81}. For our pure $8OCB$ material the
transition temperatures were $T_{IN}^o = 352.53$~K and $T_{NA}^o =
339.52$~K. The \textit{I}-\textit{N} two-phase coexistence width
was $\approx 95$~mK wide, and the \textit{N}-Sm\textit{A}
transition enthalpy was $\delta H_{NA}^o = 0.42$~J~g$^{-1}$. These
thermal features indicate that the $8OCB$ liquid-crystal used in
this study was of reasonably good quality. A summary of the
calorimetric results for pure $8OCB$ and $8OCB$+aerosil samples is
given in Table~\ref{tab:Cp-sum}.

\begin{figure}
\includegraphics[scale=0.75]{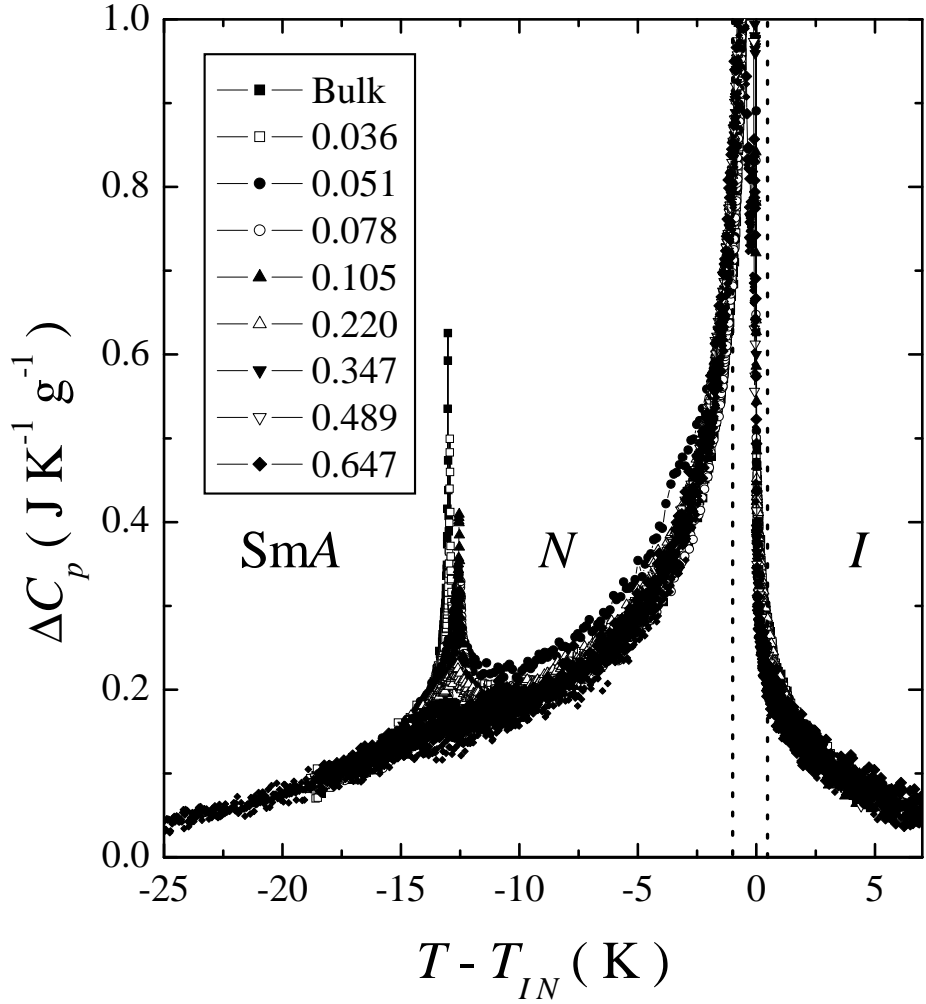}
\caption{ \label{DCPvsDTwide} Excess specific heat, $\Delta C_p$,
as a function of temperature about $T_{IN}$ for bulk $8OCB$ and
$8OCB$+aerosil samples from $\rho_S = 0.036$ to $0.647$~grams of
silica per cm$^3$ of liquid crystal. See figure inset for
definition of symbols. The vertical dashed lines indicate the
\textit{I}-\textit{N} transition region expanded in
Fig.~\ref{DCPvsDTnarr}. }
\end{figure}

In order to determine the excess heat capacity associated with the
phase transitions, an appropriate background was subtracted. The
total sample heat capacity over a wide temperature range had a
linear background, $C_p$(background), subtracted to yield
\begin{equation}
  \label{eq:DCni}
  \Delta C_p = C_p - C_p(\rm{background}) \\
\end{equation}
as the excess $C_p$ due to the \textit{I}-\textit{N} and
\textit{N}-Sm\textit{A} phase transitions. The resulting $\Delta
C_p$ data are shown for pure $8OCB$ and all $8OCB$+aerosil samples
in Fig.~\ref{DCPvsDTwide} over a wide temperature range about
$T_{IN}$, where the units are J~K$^{-1}$ per gram of
\textit{liquid crystal}. The transition temperature, $T_{IN}$, is
determined as the highest temperature where any nematic phase is
present and corresponds to the highest temperature peak in
$C_{filled}''$.

As seen in Fig.~\ref{DCPvsDTwide}, the $\Delta C_p$ values away
from the \textit{N}-Sm\textit{A} transition and the
\textit{I}+\textit{N} coexistence regions overlap with bulk
behavior independent of silica concentration. The detailed
variations of $\Delta C_p$ associated with the
\textit{N}-Sm\textit{A} transition with $\rho_S$ has been reported
previously \cite{Clegg03}. The deviations of some of the $\rho_S =
0.05$ points in the nematic phase is likely a consequence of
sample inhomogeneity. The $\Delta C_p$ ''wings'' of the
\textit{I}-\textit{N} transition are associated with short-range
fluctuations of nematic order. Given the simplification of the
nematic order parameter, the short-range fluctuations in bulk
nematics are mainly composed of thermal fluctuations of the scalar
part \textit{S}. For the $8OCB$+aerosil system, the temperature
dependence of $\Delta C_p(IN)$ being independent of $\rho_S$
suggests that thermal fluctuations of \textit{S} are independent
of disorder over the whole range of $\rho_S$ studied in this work.
The $\Delta C_p(IN)$ wing behavior shown here for $8OCB$+aerosil
is completely consistent with similar results for $8CB$+aerosil
\cite{Germano98} and low-density $8CB$+aerogel samples
\cite{Wu95}.

\begin{figure}
\includegraphics[scale=0.75]{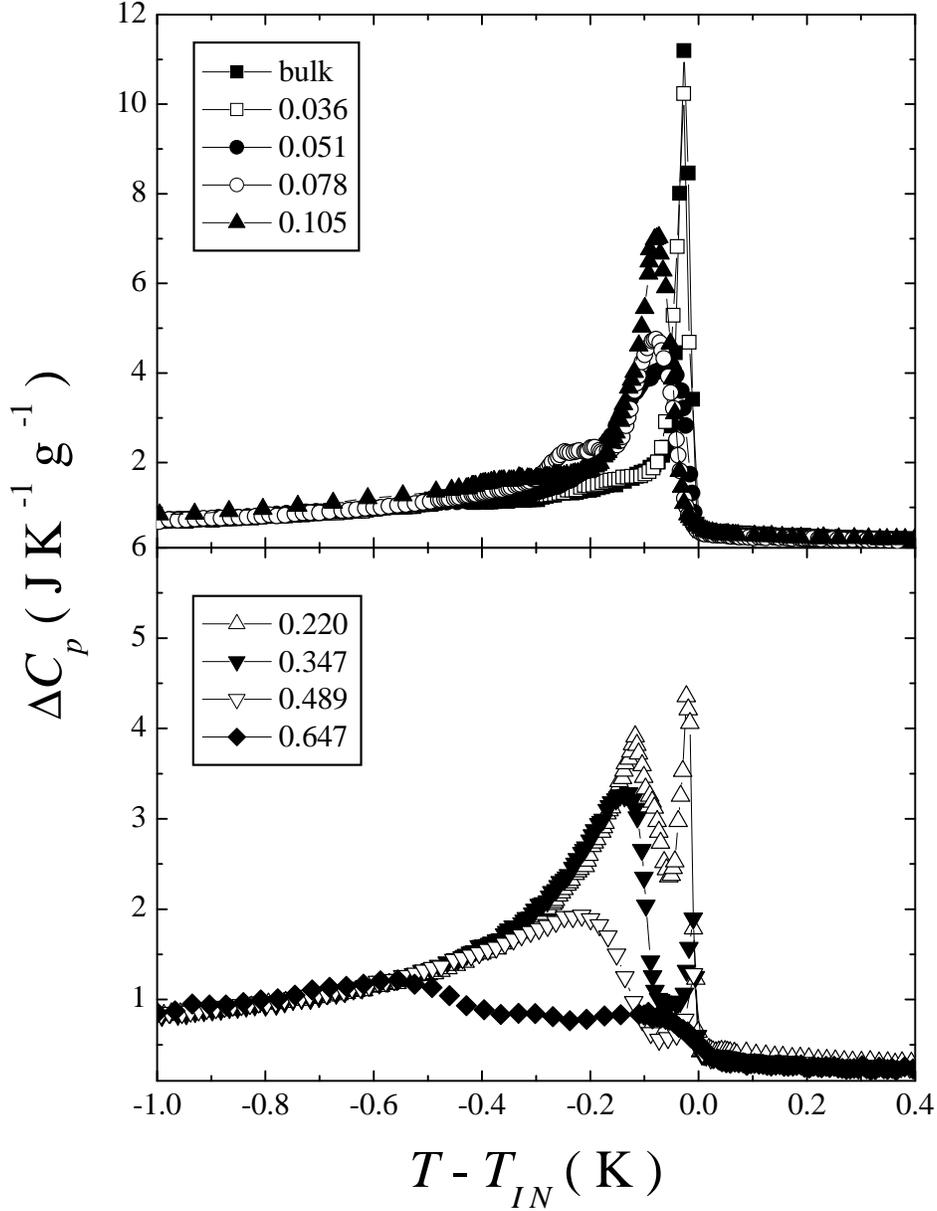}
\caption{ \label{DCPvsDTnarr} Expanded view of the excess specific
heat about the \textit{I}-\textit{N} transition as a function of
temperature. See figure insets for definition of symbols. The
samples have been separated into two groups; the upper panel
appears to indicate inhomogeneity induced broadening for samples
with $\rho_S \lesssim 0.1$; the lower panel depicts the evolution
of $\Delta C_p$ for $\rho_S > 0.1$ and shows two distinct
features, one sharp and one broad feature consistent with those
seen in $8CB$+aerosil systems \cite{Germano98}. }
\end{figure}

In stark contrast to the behavior of $\Delta C_p$ in the one-phase
regions, the two-phase coexistence region of the
\textit{I}-\textit{N} transition exhibits strong effects of silica
concentration, as shown in Fig.~\ref{DCPvsDTnarr}. From 1~K below
to 0.4~K above $T_{IN}$, the $\Delta C_p(IN)$ peaks for the pure
$8OCB$ and the $\rho_S = 0.036$ sample are essentially the same
\cite{noteBULK}. Upon increasing silica density, the peak in
$\Delta C_p(IN)$ is substantially lower in temperature relative to
the peak in $C_{filled}''$ and considerably broader than for the
bulk or the $\rho_S = 0.036$~sample. See the upper panel in
Fig.~\ref{DCPvsDTnarr}. In addition, there is a small and very
broad shoulder below the main specific heat peak, also seen in
bulk, which moves toward the main peak with increasing $\rho_S$.
The nature of this subsidiary feature is not known. It is likely,
given the similarity of the materials used here with the
$8CB$+aerosil system, that the percolation threshold for type-300
aerosil in $8OCB$ is essentially the same at $\rho_P \approx
0.018$~\cite{Germano98} and so a true gel should be present for
all samples studied in this work. This is supported by a visual
inspection of these samples holding their shape above the crystal
melting temperature. The \textit{I}-\textit{N} transition regions
shown in Fig.~\ref{DCPvsDTnarr} for $\rho_S \lesssim 0.1$ appear
to be quite sensitive to small inhomogeneities in the silica gel
dispersion and so, are grouped together. Given the dominance of a
large, broad, specific heat peak and relatively erratic transition
temperature shifts, discussed below, the effect of the silica gel
on the \textit{I}-\textit{N} phase transition is strongly
dependent on the quality of the dispersion.

Beginning with the $\rho_S = 0.22$ sample and for increasing
silica content there is a systematic variation of the excess heat
capacity, which is shown by the lower panel in
Fig.~\ref{DCPvsDTnarr}. At $\rho_S = 0.22$, a double heat capacity
feature is observed with a sharp high-temperature peak $\Delta
C_p^{HT}$ corresponding closely to (but very slightly below) a
sharp peak in $C_{filled}''$, followed at lower temperature by a
broader peak $\Delta C_p^{LT}$ also having an associated broad
peak in $C_{filled}''$. Clearly, both are first-order signatures
and they are separated by $\approx 0.1$~K. For the $\rho_S =
0.347$ and $0.489$ samples, the $\Delta C_p^{HT}$ feature remains
sharp but decreases in magnitude while the $\Delta C_p^{LT}$
feature becomes increasingly rounded and moves to lower
temperature relative to $\Delta C_p^{HT}$ by $0.15$ and $0.2$~K,
respectively. For the $\rho_S = 0.647$ sample, both heat capacity
features are rounded and separated now by $\sim 0.5$~K. Over this
entire range of silica density, the size of the $C_{filled}''$
peak decreased monotonically with increasing $\rho_S$. Such a
double \textit{I}-\textit{N} heat capacity feature was observed in
$8CB$+aerosil samples for silica concentrations up to $\rho_S
\approx 0.1$, but $\Delta C_p$ exhibited a single, rounded feature
above this density \cite{Germano98}. Only a single rounded $\Delta
C_p(IN)$ feature was observed for all $8CB$+aerogel samples
\cite{Wu95}.

\subsection{The \textit{I}-\textit{N} transition enthalpies}

The \textit{I}-\textit{N} transition enthalpy also exhibits a
dependence on aerosil concentration and can be a quantitative
measure of the strength of the transition. For a second-order (or
continuous) phase transition, the change in enthalpy through the
transition is given by
\begin{equation}
  \label{eq:DHniR}
  \delta H = \int \Delta C_{\it{p}} dT \\
\end{equation}
where the limits of integration are as wide as possible about the
heat capacity peak. However, for first-order transitions the
situation is complicated by the presence of a two-phase
coexistence region, in this work \textit{I}+\textit{N}, as well as
a latent heat $\Delta H$. The total enthalpy change through a
first-order transition is the sum of the pretransitional enthalpy
and the latent heat. In an ac-calorimetric measurement, $\Delta
C_p$ values observed in the two-phase region are artificially high
and frequency dependent due to partial phase conversion during a
$T_{ac}$ cycle. The pretransitional enthalpy $\delta H$ is
typically obtained by substituting a linearly truncated $\Delta
C_p$ behavior between the bounding points of the two-phase
coexistence region into Eq.~(\ref{eq:DHniR}), and an independent
experiment is required to determine the latent heat $\Delta H$
\cite{Germano98}. A direct integration of the observed $\Delta
C_p$ yields an effective transition enthalpy $\delta H^\ast$ and
this contains some of the latent heat contributions; thus $\delta
H < \delta H^\ast < \Delta H_{total} = \delta H + \Delta H$.

For our analysis, the observed $\Delta C_p(IN)$ was directly
integrated over a wide temperature range of $-25$~K below to
$+5$~K above $T_{IN}$ for all bulk and $8OCB$+aerosil samples
where the \textit{N}-Sm\textit{A} transition enthalpy contribution
was subtracted. This will be referred to hereafter as the
ac-enthalpy and denoted as $\delta H_{IN}^\ast$, as it represents
only a part of the total transition enthalpy. As seen in
Fig.~\ref{DCPvsDTwide}, an integration of a linearly truncated
$\Delta C_p(IN)$ in the two-phase coexistence region over a
similar range yields a pretransitional enthalpy $\delta H_{IN} =
5.13$~J~g$^{-1}$ that is independent of aerosil density.
Integration over a similar temperature range yielded a
pretransitional $\delta H_{IN}$ value of $5.43$~J~g$^{-1}$ for
$8CB$+aerosil samples, also independent of silica density
\cite{Germano98}. In addition, the integration of the imaginary
heat capacity given by Eq.~(\ref{eq:ImC}) and normalized to the LC
mass, defines an imaginary transition enthalpy, referred to as
im-enthalpy and denoted as $\delta H_{IN}''$, which is an
indicator of the first-order character of the transition. Although
$\delta H_{IN}''$ is a measure of the dispersive component of the
complex enthalpy, it is only approximately proportional to the
transition latent heat due to the fixed-$\omega$ ac-technique
employed in this work. As the silica content changes, the
two-phase conversion rate may change and so alter the
proportionality between $\delta H_{IN}''$ and $\Delta H_{IN}$;
thus a detailed frequency scan for each sample would be needed to
fully characterize the relationship. This was done for a few
samples and the frequency employed in this work is sufficiently
close to the static limit that this effect should be minimal.

\begin{figure}
\includegraphics[scale=0.75]{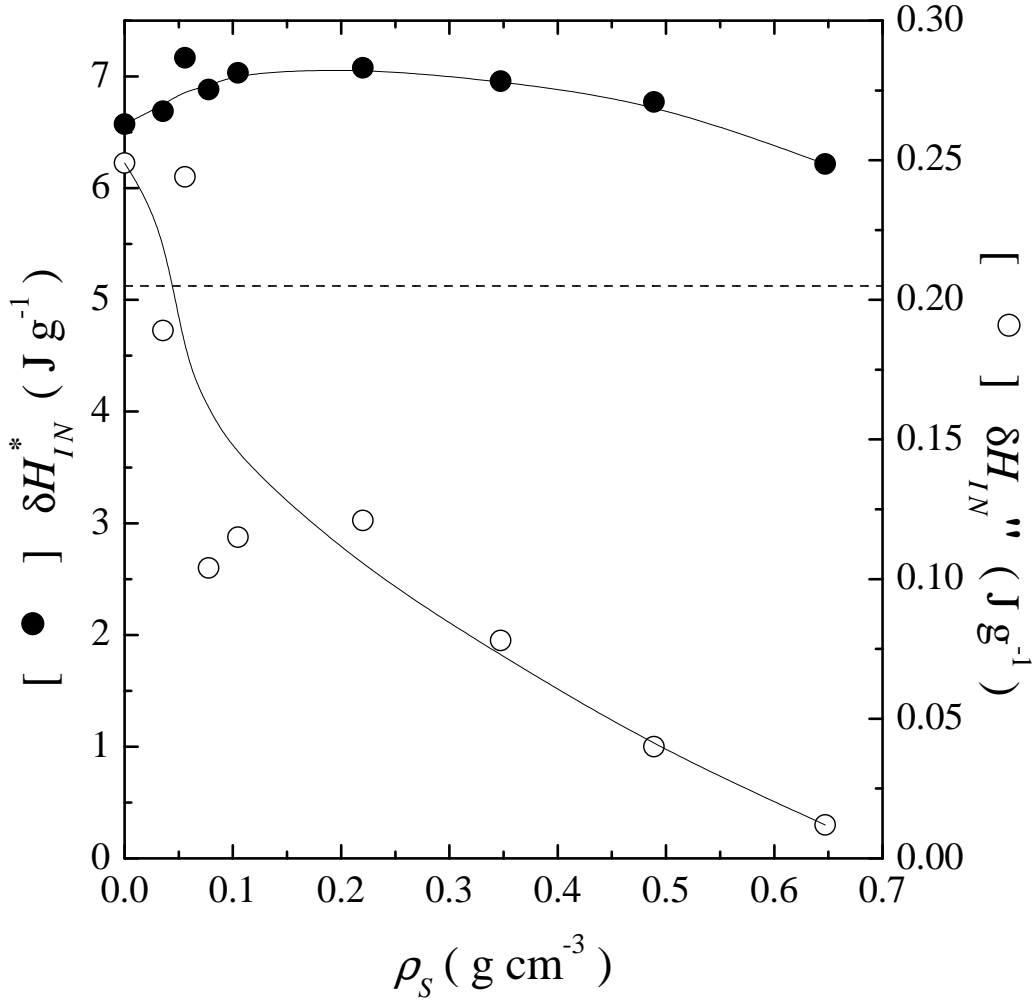}
\caption{ \label{DHNIvsRHOS} The real (solid symbol, left axis)
and imaginary (open symbol, right axis) \textit{I}-\textit{N}
transition enthalpy are shown as a function of $\rho_S$. The
effective enthalpy $\delta H_{IN}^\ast$ is weakly dependent on
silica content, indicating only minor changes to the latent heat
conversion dynamics occur relative to the ac-frequencies employed
in this work. The monotonic decrease in the imaginary component is
evidence that the first-order character of the
\textit{I}-\textit{N} transition decreases with increasing
quenched disorder. Solid lines are guides to the eye while the
horizontal dotted line represents the total pretransitional
enthalpy $\delta H_{IN} = 5.13$~J~g$^{-1}$, which is independent
of $\rho_S$. See text for details. }
\end{figure}

The results of both the ac- and im-enthalpy for $8OCB$+aerosil
samples are shown in Fig.~\ref{DHNIvsRHOS} as a function of the
silica density. There is a slight variation (first increasing for
increasing $\rho_S$ up to $0.220$ then decreasing for larger
$\rho_S$) of the ac-enthalpy due mainly to changes in $\Delta C_p$
values within the two-phase coexistence range since the heat
capacity wings away from the transition are $\rho_S$ independent
(except for $\rho_S = 0.051$, which is systematically high for $T
- T_{IN}$ from $-3$~K to $-10$~K). Given the fixed-$\omega$ aspect
of the technique, any variation observed in the ac-enthalpy in the
two-phase region can be attributable to changes in either the
dynamics or magnitude (or both) of the latent heat evolution. The
small non-monotonic variation of $\delta H_{IN}^\ast$ for
$8OCB$+aerosil samples is in contrast to the systematic decrease
of $\delta H_{IN}^\ast$ with increasing $\rho_S$ for $8CB$+aerosil
samples. This may reflect a difference in the phase conversion
dynamics between $8CB$ and $8OCB$ and how they are modified by the
presence of aerosils.

The interpretation of the im-enthalpy is more straight-forward as
it is closely related to the latent heat of the transition. With
increasing $\rho_S$, the im-enthalpy appears to monotonically
decrease to almost zero for $\rho_S = 0.647$. See
Fig.~\ref{DHNIvsRHOS}. This suggests that for the highest $\rho_S$
sample studied, the \textit{I}-\textit{N} latent heat has become
nearly zero. Similar trends were observed for $8CB$+aerosil
\cite{Germano98} where a continuous \textit{I}-\textit{N}
transition is estimated to occur near $\rho_S \approx 0.8$. Also,
a nearly continuous \textit{I}-\textit{N} transition was reported
for $7CB$+aerosil for silica densities near $\rho_S \approx 1$
\cite{Jamee02}. The above observations are consistent with the
general view that with increasing QRD, first-order transitions are
driven continuous \cite{Imry79}.

\subsection{Transition temperatures and crystallization}

\begin{figure}
\includegraphics[scale=0.75]{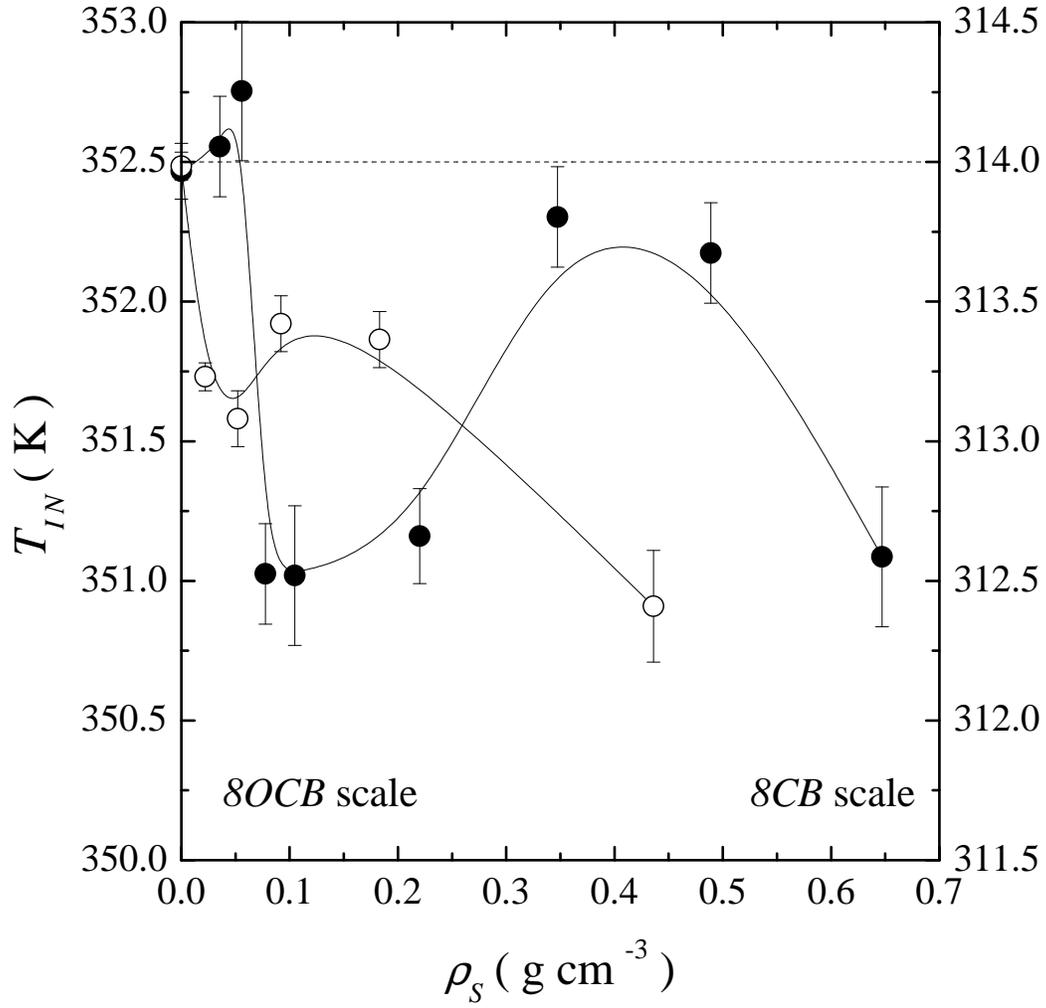}
\caption{ \label{TNIvsRHOS} Dependence on $\rho_S$ of the
\textit{I}-\textit{N} transition temperature, $T_{IN}$, for
$8OCB$+aerosil (solid circles and left axis, $T_{IN}^o =
352.47$~K) and $8CB$+aerosil (open circles and right axis,
$T_{IN}^o = 313.99$~K) samples. Data for $8CB$+aerosil samples
taken from Ref.~\cite{Germano98}. Note that both the left and
right axes span $3$~K in temperature. The solid lines are guides
to the eye. }
\end{figure}

The \textit{I}-\textit{N} transition temperature, defined here as
the peak in $C_{filled}''$ for the highest temperature feature,
for the $8OCB$+aerosil samples as well as those for the
$8CB$+aerosil system taken from Ref.~\cite{Germano98} are shown in
Fig.~\ref{TNIvsRHOS} as a function of silica density. For the
$8OCB$+aerosil system, $T_{IN}$ is essentially unchanged up to
$\rho_S = 0.051$ then decreases sharply by $\sim 1.5$~K at $\rho_S
= 0.078$, $0.105$, and $0.220$. It then rises strongly for $\rho_S
= 0.347$, nearly recovering the bulk value. Upon further increase
in $\rho_S$, $T_{IN}$ decreases monotonically (with a concave
downward character) until it is again about $\sim 1.5$~K below
$T_{IN}^o$ for the $\rho_S = 0.647$ sample. The non-monotonic
evolution of $T_{IN}$ with silica content for the $8OCB$+aerosil
system is similar to that seen in the $8CB$+aerosil system,
suggesting that the initial depression of $T_{IN}$, recovery, then
continued depression is a general phenomena of quenched random
disorder on nematics while the specific $\rho_S$ dependence is
liquid crystal material dependent. Over this same range in silica
density, the width of the two-phase coexistence region $\delta
T_{IN}$ also has a non-monotonic dependence on $\rho_S$, as seen
in Table~\ref{tab:Cp-sum}. However, $\delta T_{IN}$ is sensitive
to local inhomogeneities of the aerosil dispersion that may
account for its variation when $\rho_S < 0.1$. Beginning at
$\rho_S = 0.105$, $\delta T_{IN}$ increases monotonically by a
factor of $\sim 6.7$ while $\rho_S$ increases by a factor of $\sim
6$. The observed broadening of the two-phase coexistence width in
nearly direct proportion with increasing QRD is generally
consistent with the behavior of first-order transitions with
quenched disorder \cite{Imry79}.

\begin{figure}
\includegraphics[scale=0.75]{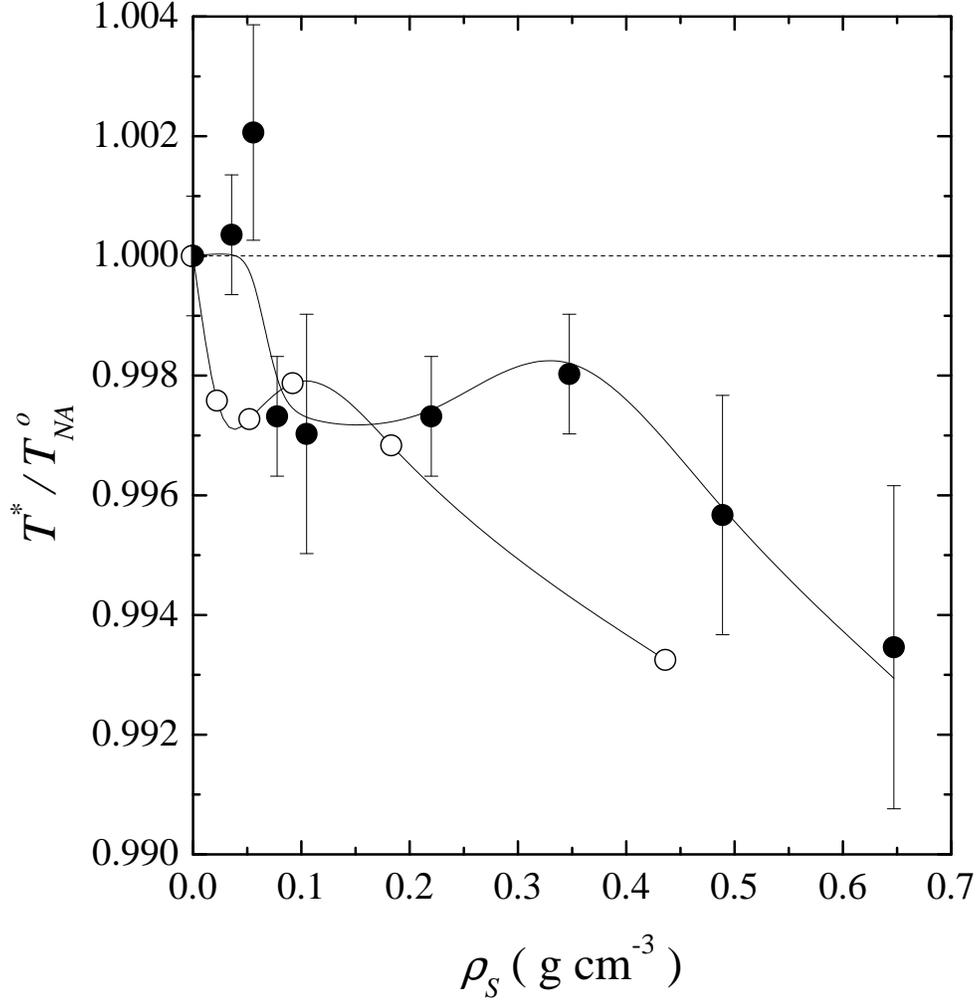}
\caption{ \label{TNAvsRHOS} Dependence on $\rho_S$ of the
pseudo-transition \textit{N}-Sm\textit{A} temperature, $T^\ast$
scaled by the bulk value for $8OCB$+aerosil (solid circles,
$T_{NA}^o = 339.52$~K) and $8CB$+aerosil (open circles, $T_{NA}^o
= 306.97$~K) samples. Data for $8CB$+aerosil samples taken from
Ref.~\cite{Germano98}. The solid lines are guides to the eye. }
\end{figure}

The \textit{N}-Sm\textit{A} pseudo-transition temperatures
$T^\ast$ scaled by the bulk transition temperature $T_{NA}^o$ for
$8OCB$+aerosil and $8CB$+aerosil systems are shown in
Fig.~\ref{TNAvsRHOS}. The pattern of fractional changes in the
$T^\ast$ is essentially the same for both LC+aerosil systems with
an initial rapid depression, recovery, then more gradual decrease
with a total change of less than $1\%$ from $T_{NA}^o$. The
primary difference with the $8CB$+aerosil system is that this
behavior is ''stretched'' in $\rho_S$ for the $8OCB$+aerosil
samples. This is consistent with the evolution seen of $T_{IN}$
shown in Fig.~\ref{TNIvsRHOS} and described above.

\begin{figure}
\includegraphics[scale=0.75]{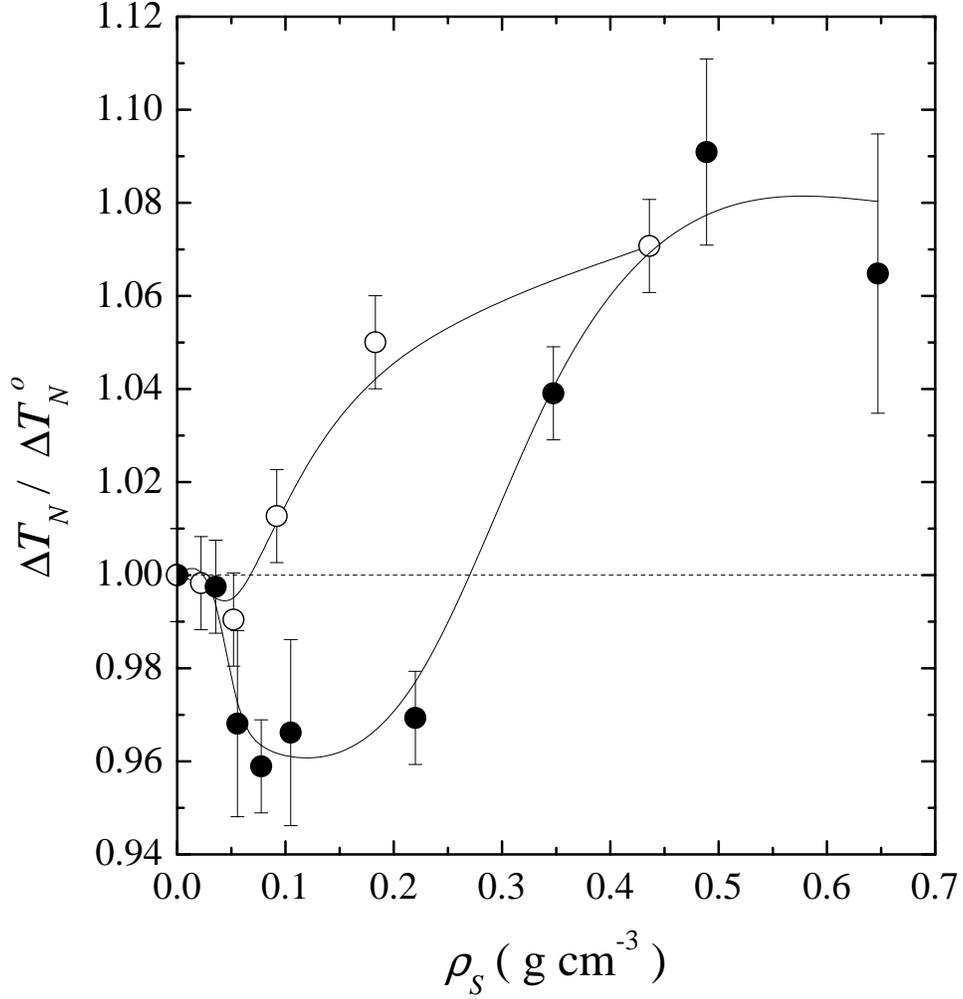}
\caption{ \label{NEMvsRHOS} The nematic phase temperature range
$\Delta T_N = T_{IN} - T^\ast$ scaled by the bulk value for
$8OCB$+aerosil (solid circles, $\Delta T_N^o = 12.95$~K) and
$8CB$+aerosil (open circles, $\Delta T_N^o = 7.01$~K) samples.
Data for $8CB$+aerosil taken from Ref.~\cite{Germano98}. The solid
lines are guides to the eye. }
\end{figure}

The nematic phase temperature range, $\Delta T_N = T_{IN} -
T^\ast$, normalized by the bulk nematic range $\Delta T_N^o =
T_{IN}^o - T_{NA}^o$, is shown in Fig.~\ref{NEMvsRHOS}. While the
individual transition temperature changes reflect the absolute
stability limit of the nematic and smectic phases, $\Delta T_N$
reflects the relative stability of both phases. For $8CB$+aerosil,
a decrease of $\sim 1\%$ in $\Delta T_N$ was seen up to $\rho_S
\approx 0.1$, corresponding to the the local maximum of
$T_{IN}(\rho_S)$ and $T^\ast(\rho_S)$. This was originally thought
to be scatter in the data of Ref.~\cite{Germano98}. For
$8OCB$+aerosil samples, a similar and far more pronounced $4\%$
decrease in $\Delta T_N$ is seen from $\rho_S = 0.051$ to $0.220$.
This decrease in $\Delta T_N$ reflects a greater depression of
$T_{IN}$ than $T^\ast$ and indicates that in this range of silica
density, the disorder primarily effects nematic (orientational)
ordering. Upon further increasing $\rho_S$, the nematic range
begins to increase and appears to saturate at an $8\%$ increase
similar to that seen in the $8CB$+aerosil system. This growth in
the nematic range occurs because of the greater suppression of
$T^\ast$ relative to $T_{IN}$ and so, reflects that above $\rho_S
\gtrsim 0.2$, the effect of the silica gel is to mainly disorder
smectic (1D-translational) ordering.

\begin{figure}
\includegraphics[scale=0.75]{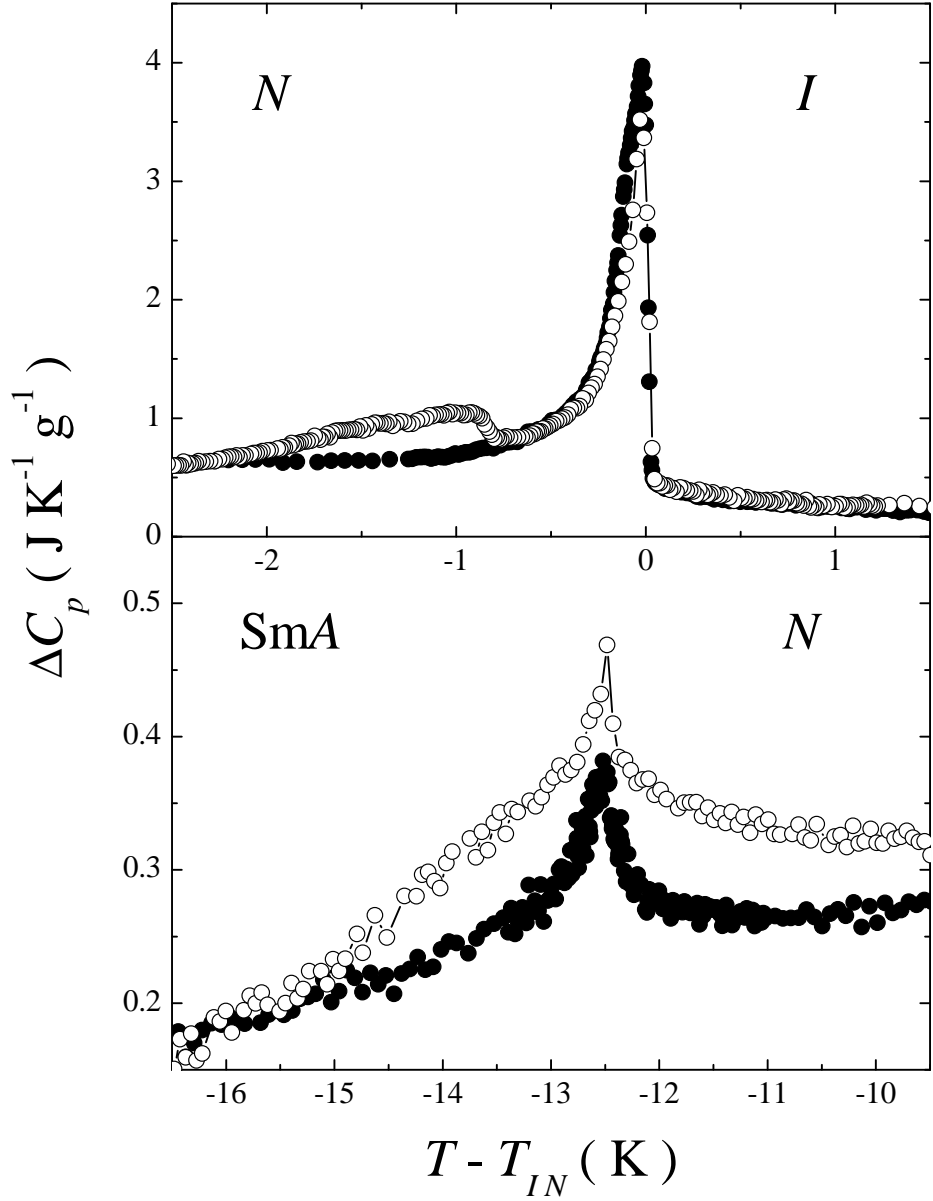}
\caption{ \label{DCP050vsDT} Behavior of the \textit{I}-\textit{N}
and \textit{N}-Sm\textit{A} excess specific heat of the $\rho_S =
0.051$ sample as a function of temperature relative to $T_{IN}$
before (solid circles, $T_{IN} = 352.76$~K ) and after (open
circles, $T_{IN} = 352.07$~K ) sample crystallization. Both are
heating scans made under identical ac-calorimetry conditions. Note
the excess enthalpy for the \textit{N}-Sm\textit{A} transition and
the second feature near the \textit{I}-\textit{N} transition as
well as a $\sim 0.7$~K shift \textit{downward} of $T_{IN}$
observed after crystallization. See text for details.}
\end{figure}

\begin{figure}
\includegraphics[scale=0.75]{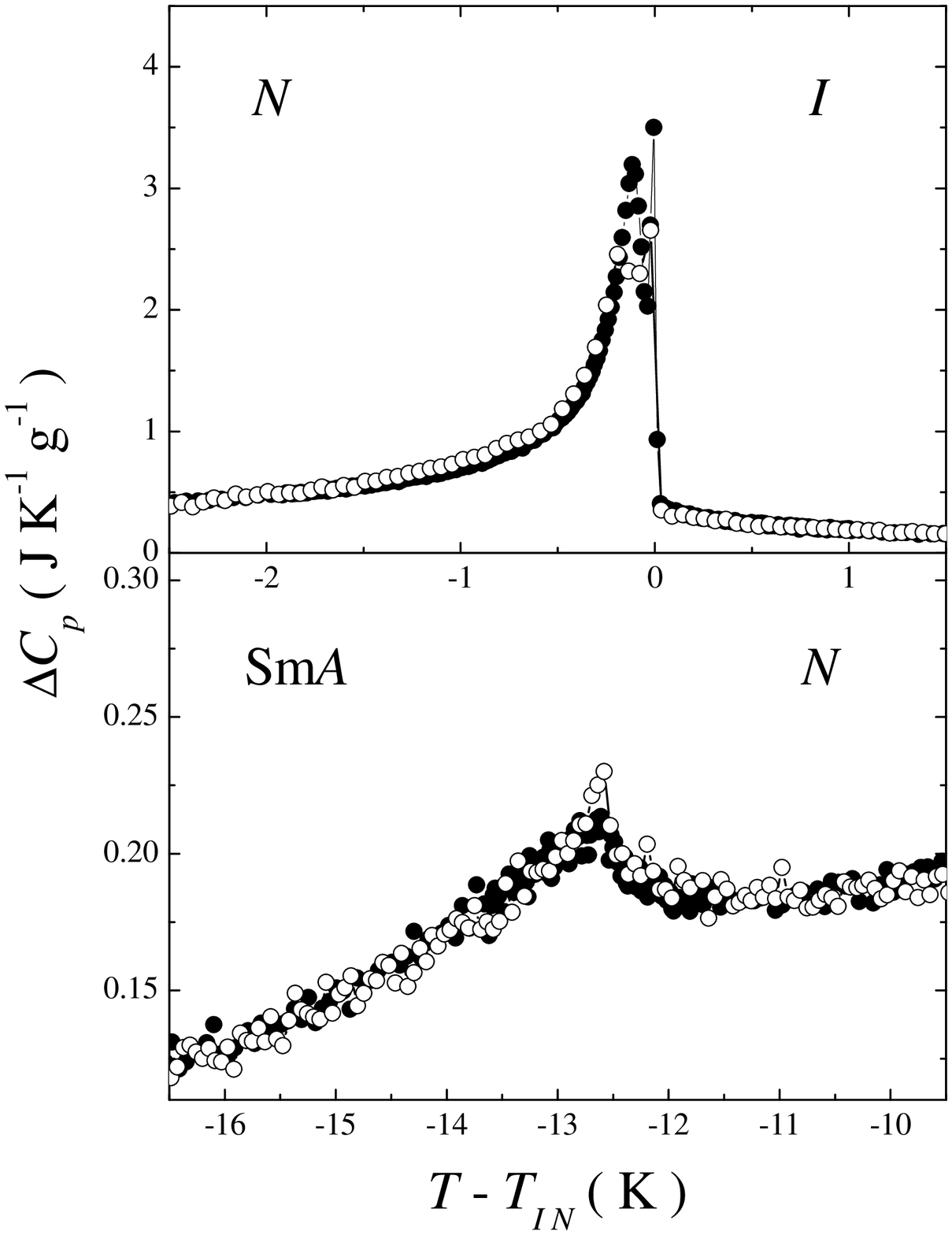}
\caption{ \label{DCP200vsDT} Behavior of the \textit{I}-\textit{N}
and \textit{N}-Sm\textit{A} excess specific heat of the $\rho_S =
0.220$ sample as a function of temperature about $T_{IN}$ before
(solid circles, $T_{IN} = 351.02$~K ) and after (open circles,
$T_{IN} = 351.23$~K ) sample crystallization. The after scan used
the same ac-input power but a faster temperature scan rate than
the scan before crystallization; both are heating scans. Note the
nearly perfect reproducibility of $\Delta C_p$ with a $T_{IN}$
shift \textit{upward} of $\sim 0.2$~K observed after
crystallization. See text for details.}
\end{figure}

The calorimetric results on the \textit{I}-\textit{N} transition
temperature described above suggest the importance of sample
homogeneity. As a test of the fragility of the silica gel, heat
capacity scans were performed on a low and high density
$8OCB$+aerosil sample immediately before and after crystallization
of the LC. Such a thermal cycle for the $\rho_S = 0.051$ sample is
shown in Fig.~\ref{DCP050vsDT} and for the $\rho_S = 0.220$ sample
in Fig.~\ref{DCP200vsDT} as a function of $\Delta T = T - T_{IN}$
in order to suppress hysteresis effects of the
\textit{I}-\textit{N} transition. The effect of crystallization on
the $\rho_S = 0.051$ sample is striking, revealing significant
distortion of the $\Delta C_p$ signature at both the
\textit{I}-\textit{N} and \textit{N}-Sm\textit{A} transitions. The
appearance of an additional broad $\Delta C_p$ feature beginning
$\sim 0.9$~K below $T_{IN}$ as well as a broadened feature over
the \textit{N}-Sm\textit{A} transition region after
crystallization may indicate increased sample inhomogeneity,
presumably caused by the local expulsion of silica particles as LC
crystallites form. However, there are two puzzling aspects; (1)
the shift in $T_{IN}$ is \textit{downward} by $\sim 0.7$~K and (2)
the specific distortion seen in $\Delta C_p$ about $T^\ast$
reveals an \textit{increased} \textit{N}-Sm\textit{A} transition
enthalpy. The first aspect is counter-intuitive as the expulsion
of impurities upon crystallization should have moved the system
closer to bulk behavior by increasing the size of pure LC domains
(regions where no silica is present). The second aspect is
particularly puzzling as the \textit{N}-Sm\textit{A} transition
enthalpy after crystallization surpasses the
\textit{N}-Sm\textit{A} transition enthalpy for \textit{bulk}
$8OCB$ ($\Delta H_{NA}(before) = 0.334$, $\Delta H_{NA}(after) =
0.65$, and $\Delta H_{NA}(bulk) = 0.42$ all given in units of
J~g$^{-1}$). Visual inspection of the sample immediately after the
first crystallization revealed no obvious inhomogeneities. A
recent Raman spectroscopy study of $8CB$ crystallizing within a
gel matrix provides evidence of new solid and semi-solid phase at
low temperature \cite{fehr03}. The additional enthalpy observed
here could indicate new solid phases for $8OCB$+aerosil samples.
These results are only observed upon initial crystallization of a
freshly dispersed $8OCB$+aerosil sample. Bulk $8OCB$ behavior is
eventually approached upon repeated thermal cycling through the
crystallization transition. Clearly, at these low silica
densities, the silica gel is, at least locally, quite fragile and
short range restructuring of the gel can strongly affect the
liquid-crystal.

The results of initial cycling through the crystallization
transition for the $\rho_S = 0.220$ sample shown in
Fig.~\ref{DCP200vsDT} do not reveal any significant changes. Aside
from a very small, sharp, additional heat capacity feature at the
\textit{N}-Sm\textit{A} transition, the $\Delta C_p$ curves are
almost perfectly reproducible. The shift in $T_{IN}$ after the
initial crystallization is small and upward by $\sim 0.2$~K, as
expected by the expulsion of impurities. This indicates that for
higher silica densities, the gel is robust and well behaved. Note
that there is no appreciable change in the nematic range, as seen
in both Fig.~\ref{DCP050vsDT} and \ref{DCP200vsDT}.


\section{DISCUSSIONS AND CONCLUSIONS}
\label{sec:disc}

Results have been presented from high-resolution ac-calorimetric
experiments on $8OCB$+aerosil dispersions with emphasis on the
weakly first-order \textit{I}-\textit{N} phase transition. These
results for $8OCB$+aerosil dispersions have been compared with
existing results for $8CB$+aerosil dispersions
\cite{Germano98,Germano03} and reveal new aspects of the effect of
quenched random disorder on liquid crystal phase transitions. In
particular, these two LC+aerosil systems are very similar except
for the relative elasticity of the LC material. The material
$8OCB$ is elastically stiffer, having an effective (single)
nematic elastic constant $K_N$ larger by approximately $20\%$,
than $8CB$, and this is reflected by the higher transition
temperatures for the nematic, smectic, and crystal phases. Thus,
aspects that are LC material dependent and those that are general
to quenched random disorder can be distinguished.

From the very good overlap of the $\Delta C_p$ wings away from the
\textit{I}-\textit{N} two-phase coexistence region and from direct
NMR studies on $8CB$+aerosil \cite{Jin01} a general feature of
LC+aerosils is that the magnitude of the nematic order, $S$, is
essentially the same as in the bulk LC. Thus, the main effect of
QRD is on the director structure with the elasticity of the LC and
the kinetics of the ordered phase growth as likely important
factors. This is supported by the differences seen between
$8OCB$+aerosil and $8CB$+aerosil systems; in particular, the
different $\rho_S$ dependence is likely connected to the
difference in elasticity for the two liquid-crystals.

The effect of quenched random disorder on first-order phase
transitions is substantially different than that on continuous
phase transitions. First-order transitions have an additional
energy penalty for the formation of interfaces between coexisting
phases, which complicates random-field type theoretical
approaches. In the classical treatment, first developed by Imry
and Wortis \cite{Imry79}, the QRD effect on first-order
transitions is that a quenched random-field creates domains having
a randomly shifted transition temperature. This would have the
effect of smearing the overall transition, a monotonic decrease in
the transition latent heat and temperature with increasing QRD,
and a low-temperature phase possessing only short-range order for
arbitrarily weak QRD. These predictions are generally consistent
with the behavior of nematic in aerogels \cite{Wu95}. For nematics
in aerosils, the transition latent heat appears to decrease and
the width of the coexistence region to increase monotonically with
increasing disorder as well as the transition becoming apparently
continuous for high disorder strength. These features are
generally consistent with the classical picture. However, the
character of the transition and the non-monotonic transition
temperature shifts do not appear to be consistent with this view.
More strikingly, the \textit{I}-\textit{N} transition in aerosils
in some range of $\rho_S$, appears to proceed via two transitions.
This could only occur in the classical view if a bimodal
distribution of the random-field variance $\langle h^2\rangle$,
connected to a bimodal distribution in $\rho_S$, is present. This
is not supported by SAXS studies, which revealed the fractal-like
nature of the aerosil gel structure \cite{Germano98}, nor the
behavior at the \textit{N}-Sm\textit{A} transition
\cite{Haga97a,Haga97b,Germano98}. However, nematics are very
"soft" materials and the QRD imposed by aerosils appears to be
much weaker than that of aerogels, thus elasticity of the LC (and
possibly the gel) can play an important role.

In LC+aerosil systems, $\langle h^2\rangle$ is thought to depend
on the given concentration of silica and interaction with the LC,
whereas the LC elasticity $K_N$ is strongly temperature dependent
(being proportional to $S^2$) near and below $T_{IN}$. Since the
aerosil gel is thixotropic and formed in the isotropic phase, any
high energy strains or deformations that may exist are likely
quickly annealed, the anisotropy of the gel should be fixed and
essentially zero, especially for rigid gel structures like
aerogels. However, the disordering nature of the gels may evolve
with thermal history of the LC in aerosil gels. As seen by the
result of cycling through crystallization presented here as well
as the DNMR \cite{Jin01} and electro-optical
\cite{Bellini00,Bellini02} studies, the gel can be compliant with
respect to distortions in the director structure for a range of
silica densities. Note that the quantity $T_{IN}$ differs by at
most 10\% between $8CB$ and $8OCB$ thus the energy scales are
similar while the twist elastic constant differs by 36\%.

The features described above suggests a possible physical scenario
for the origin of the double $\Delta C_p$ feature at the
\textit{I}-\textit{N} (or any ''soft'' first-order
\cite{noteLCSIL}) transition in LC+aerosil systems. As the nematic
elastic constant strongly increases with decreasing temperature
for $T \lesssim T_{IN}$, a ''skin'' of low nematic order (due to
the undulations of the aerosil strands) may coat the silica
strands. The thickness of this paranematic boundary layer would be
strongly temperature dependent, shrinking with decreasing $T$
below $T_{IN}$. The presence of such layer a would serve to
partially decouple the disordering (or field) effect of the silica
gel from the void nematic (acting as a kind of ''lubricant'').
Once the layer thickness reaches its minimum value (roughly
equivalent to a molecular length) the elasticity of the void
nematic becomes strongly coupled to that of the aerosil gel. This
would effectively increase, for a given $\rho_S$, the disorder
strength.

A consequence of this speculation for a first-order transition
induced change in disorder strength (through the onset of coupling
between the director fluctuations and gel) would be the alteration
of the gel dynamics (i.e., vibrational modes). This would be
consistent with large changes in the relaxation times of aerosil
gels observed by dynamic x-ray studies on $8CB$+aerosil near
$T_{IN}$ \cite{Retsch02}. In addition, this coupling should dampen
director fluctuations and could account for the variation in the
critical behavior seen at the \textit{N}-Sm\textit{A} transition
for LC+aerosil samples. Another consequence of this view is that
the director correlation length $\xi_{\hat n}$ (the relevant
aspect of nematic order) would jump to a large isotropic value at
the first transition and upon further cooling cross a second
transition into a more strongly disordered state having a
\textit{smaller} correlation length. The isotropic nature of
$\xi_{\hat n}$ and the final SRO state of the nematic with QRD are
consistent with recent optical studies \cite{Bellini02} and the
detailed evolution of $\xi_{\hat n}$ through the two transitions
is the subject of current optical and calorimetric study
\cite{noteBELLINI}.

The different silica density dependence of $T_{IN}$ and the
temperature distance between the two $C_p$ peaks ($\delta T_{2p}$)
between $8OCB$+aerosil and $8CB$+aerosil would also be consistent
with the difference in the nematic elasticity of the two
liquid-crystals. A stiffer silica gel (higher $\rho_S$) would be
required to influence a stiffer LC, thereby stretching the shift
in $T_{IN}$ with respect to $\rho_S$ as seen between the
$8OCB$+aerosil and $8CB$+aerosil systems. The $\rho_S$ dependence
in the high-density regime of $\delta T_{2p}$ for $8OCB$+aerosil
(see Table~\ref{tab:Cp-sum}) compared to the nearly constant
$\delta T_{2p} \sim 0.1$~K seen only in the low density regime of
$8CB$+aerosil \cite{Germano98} would be compatible with a
transition induced increase in the variance of the gel disorder
strength. The liquid-crystal $8CB$ being much softer would only be
able to stress a very weak (low $\rho_S$) gel while the much
stiffer $8OCB$ would be able to distort a wider range of gels.
Since the $\rho_S$ dependence of the critical behavior for the
\textit{N}-Sm\textit{A} transition is quite similar between
$8CB$+aerosil and $8OCB$+aerosil \cite{Clegg03}, these
observations suggest that the effects observed at the
\textit{I}-\textit{N} transition are not directly connected to
those at the \textit{N}-Sm\textit{A}.

Finally, the unexpected behavior of low silica density sample when
initially cycled through the crystallization transition is not
fully understood. The expulsion of silica impurities by the
strongly first-order crystallization transition seems to lead to a
\textit{more} disordered system. One possibility to explain this
phenomena is that the initial crystallization causes the expulsion
of the silica particles locally and transforms the flexible
fractal structure into a more rigid, foam-like, gel. The
depression of $T_{IN}$ is then a consequence of the greater
elastic distortions imposed by the new gel arrangement. Repeated
cycling through crystallization would continue to expel
impurities, eventually destroying the foam-like structure. Once
the silica has been compacted sufficiently, percolation is no
longer possible and free-floating silica particles would represent
an annealed disorder. The gels formed in the high silica density
samples are more robust and would not be expected to change
significantly when the LC crystallize. This view can be directly
tested with a detailed structural study by small-angle x-ray
scattering where the thermal history is carefully controlled. The
increase in the enthalpy of the \textit{N}-Sm\textit{A} transition
after the initial crystallization to a value greater than the bulk
LC value remains a puzzle. Note that the speculations presented
here for the double \textit{I}-\textit{N} transition peaks and the
unusual hysteresis behavior are intended to motivate future
experimental and theoretical studies.


\begin{acknowledgments}

The authors wish to thank C. W. Garland, Robert Leheny, and
Tommaso Bellini for many helpful and useful discussions. Funding
in Toronto was provided by the Natural Science and Engineering
Research Council of Canada, and the work at WPI was supported by
the NSF under the NSF-CAREER award DMR-0092786.

\end{acknowledgments}


\begin{table*}
\caption{ \label{tab:Cp-sum} Summary of the calorimetric results
for $8OCB$+aerosil samples.  Shown are the sample density
($\rho_S$ in grams of aerosil per cm$^3$ of $8OCB$) as well as the
\textit{I}-\textit{N} ($T_{IN}$) and the \textit{N}-Sm\textit{A}
($T^\ast$) phase transition temperatures, the nematic range
($\Delta T_{N} = T_{IN} - T^\ast$), the width of the
\textit{I}-\textit{N} coexistence region ($\delta T_{IN}$), and
the difference in temperature between the two $\Delta C_p(IN)$
peaks ($\delta T_{2p}$) all in kelvins. The \textit{I}-\textit{N}
transition ac-enthalpy ($\delta H_{IN}^\ast$) and the integrated
imaginary specific heat ($\delta H_{IN}''$) in J~g$^{-1}$ are also
tabulated. }
\begin{ruledtabular}
\begin{tabular}{@{\extracolsep{20pt}}llllllll}
  $\rho_S$ & $T_{IN}$ & $T^\ast$ & $\Delta T_{N}$ & $\delta T_{IN}$ & $\delta T_{2p}$ & $\delta H_{IN}^\ast$ & $\delta H_{IN}''$ \\
 \hline
  $0$      & $352.47$ & $339.52$ & $12.95$ & $0.10$ & $-$    & $6.57$ & $0.25$ \\
  $0.036$  & $352.56$ & $339.64$ & $12.92$ & $0.11$ & $-$    & $6.69$ & $0.19$ \\
  $0.051$  & $352.75$ & $340.22$ & $12.56$ & $0.32$ & $-$    & $7.17$ & $0.24$ \\
  $0.078$  & $351.03$ & $338.61$ & $12.42$ & $0.27$ & $-$    & $6.88$ & $0.10$ \\
  $0.105$  & $351.02$ & $338.51$ & $12.51$ & $0.18$ & $-$    & $7.03$ & $0.12$ \\
  $0.220$  & $351.16$ & $338.61$ & $12.55$ & $0.27$ & $0.09$ & $7.08$ & $0.12$ \\
  $0.347$  & $352.30$ & $338.85$ & $13.45$ & $0.46$ & $0.11$ & $6.96$ & $0.08$ \\
  $0.489$  & $352.17$ & $338.05$ & $14.12$ & $0.70$ & $0.22$ & $6.77$ & $0.04$ \\
  $0.647$  & $351.09$ & $337.30$ & $13.79$ & $1.20$ & $0.54$ & $6.22$ & $0.01$ \\
\end{tabular}
\end{ruledtabular}
\end{table*}


\begin{thebibliography}{32}
\bibitem[{not({\natexlab{a}})}]{noteBULK}
\bibinfo{note}{A very small feature whose exact origin is unclear but may
  reflect an annealed impurity is seen in bulk at $\Delta T = T - T_{NI}
  \approx -0.5$~K and shifts upward relative to $T_{NI}$ for the $\rho_S =
  0.036$~sample.}
\end{thebibliography}
\end{document}